\documentclass[aps,prl,twocolumn,showpacs,showkeys]{revtex4}

\usepackage{amsmath,amssymb,epsfig,amsfonts}
\usepackage{graphicx}

\begin{document}

\title{Storage and Adiabatic Cooling of Polar Molecules in a Microstructured Trap}
\author{B.\,G.\,U.~Englert}
\author{M.~Mielenz}
\author{C.~Sommer}
\author{J.~Bayerl}
\author{M.~Motsch}
\thanks{Present address: Laboratorium f\"ur Physikalische Chemie, ETH Z\"urich, CH-8093, Switzerland}
\author{P.\,W.\,H. Pinkse}
\thanks{Present address: MESA+ Institute for Nanotechnology, University of Twente, 7500AE, The Netherlands}
\author{G.~Rempe}
\author{M.~Zeppenfeld}
\email{martin.zeppenfeld@mpq.mpg.de}

\affiliation{Max-Planck-Institut f\"ur Quantenoptik, Hans-Kopfermann-Str. 1, D-85748 Garching, Germany}

\begin{abstract}
We present a versatile electric trap for the exploration of a wide range of quantum phenomena in the interaction between polar molecules. The trap combines tunable fields, homogeneous over most of the trap volume, with steep gradient fields at the trap boundary. An initial sample of up to $10^8$ CH$_3$F molecules is trapped for as long as 60 seconds, with a 1/e storage time of 12 seconds. Adiabatic cooling down to $120\,$mK is achieved by slowly expanding the trap volume. The trap combines all ingredients for opto-electrical cooling, which, together with the extraordinarily long storage times, brings field-controlled quantum-mechanical collision and reaction experiments within reach.
\end{abstract}



\maketitle
Polar molecules with their numerous internal degrees of freedom and strong long-range interactions are ideal systems for the investigation of fundamental phenomena of cold and ultracold matter. They allow unique approaches to quantum computation~\cite{DeMille02,Andre06}, can condense to new quantum phases~\cite{Goral02,Micheli06}, and are promising candidates for precision tests of fundamental symmetries~\cite{Hinds97,Vutha10,Hudson11}. Moreover, novel quantum-mechanical collision and reaction channels are predicted for cold molecules~\cite{Krems08}. Here, field-induced alignment~\cite{Miranda11} and field-sensitive collision resonances~\cite{Avdeenkov02} allow the study of controlled chemistry~\cite{Tscherbul08}. The experimental exploration of such interaction-induced phenomena requires dense and cold molecular gases. To produce these, electric trapping techniques provide a key advantage by combining long interaction times and good localization with deep confinement of the molecules~\cite{Meerakker05a,Rieger05,Kleinert07,Hogan09}. However, the huge Stark broadening induced by the inhomogeneous trapping fields, on the order of $10\,$GHz for the achievable molecular temperatures, precludes the application of traps in precision spectroscopy and collision experiments. Therefore, new molecular cooling and trapping techniques have to be developed for all investigations involving narrow resonances. Moreover, electric trap lifetimes are so far limited to around a second, not long enough to observe molecular collisions with the attainable densities for non-alkali molecules. 

Here we report on the experimental realization of a novel electric trap featuring several key innovations. Specifically, polar molecules are trapped in a box-like potential where variable homogeneous electric fields can be applied to a large fraction of the trap volume. High trapping fields exist only at the trap boundary. This allows electric-field sensitive collision resonances and optical transitions to be addressed with strongly suppressed Stark broadening. Molecules are stored as long as a minute with a 1/e time of 12 seconds, about an order of magnitude longer than in any other electric trapping experiment reported to date. The trap is continuously loaded by a guided beam of cold molecules and is closed when sufficient molecules are stored. The trapped molecules are then cooled by adiabatic expansion, making use of a unique feature of our trap, namely the subdivision into two trap regions where homogeneous electric fields can be applied independently. This expansion occurs along one direction but is shown to cool in all three dimensions due to mixing of all motional degrees of freedom. The observed cooling is limited by the trap dimensions, but large temperature reductions are expected for opto-electrical cooling~\cite{Zeppenfeld09}, a general Sisyphus-type cooling scheme for polar molecules which can be ideally implemented in our trap. The exceptional versatility and outstanding performance of the trap makes it an ideal toolbox with a promising application potential in polar molecule experiments.

\begin{figure}[t]
\centering
\includegraphics[width=0.48\textwidth]{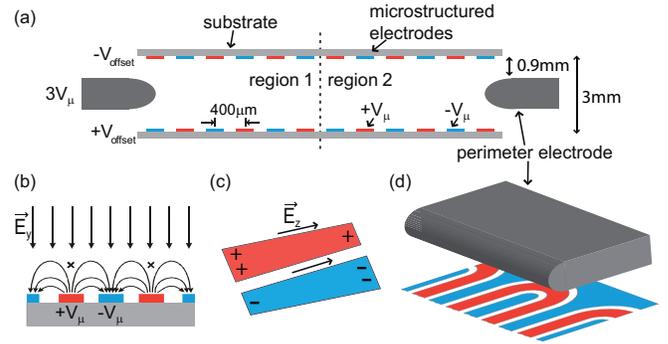}
\caption{(color online). 
(a) Side view of the electric trap (not to scale). The trap consists of a high-voltage perimeter electrode and capacitor plates with microstructured surface electrodes. Trap dimensions are $4\,{\rm cm}\times 2\,{\rm cm} \times 3\,{\rm mm}$. (b)-(d) Details of the electric-field configuration and microstructure electrode design as discussed in the text.}
\label{fig:trap}
\end{figure}

A schematic of the trap is presented in Fig.~\ref{fig:trap}(a). Two parallel capacitor plates produce tunable homogeneous electric fields in a large fraction of the two trap regions~1 and~2. To prevent molecules from colliding with the plate surfaces, the capacitor plates consist of a planar array of equidistant ($400\,\rm{\mu m}$) microstructured electrodes deposited on a  glass substrate. Applying opposite-polarity voltages $\pm V_{\mu}$ to adjacent electrode stripes creates large repelling electric fields near the plate surfaces which exponentially decay away from the plates~\cite{Wark92,Schulz04}. Between the plates, a background field is produced by applying additional offset voltages $\pm V_{\rm offset}$ to the two plates. Transverse confinement of the molecules is achieved by a high-voltage electrode between the plates that surrounds the perimeter of the trap. 
With this design we achieve a uniform confining electric field strength of up to $60\,\rm{\frac{kV}{cm}}$.

To avoid severe trap losses, attention must be paid to the details of the electric fields. As shown in Fig~\ref{fig:trap}(b), the interference of the microstructure field with a homogeneous offset field gives rise to a zero electric field (indicated by crosses in Fig.~\ref{fig:trap}(b)) above every second microstructure electrode. These zeros cause trap losses in two ways: First, molecules are likely to undergo non-adiabatic transitions, so-called Majorana flips, to states that are no longer trapped~\cite{Kirste09}. Second, these zeros continue underneath the perimeter electrode, allowing molecules to leak out of the trap volume. To reduce Majorana flips, the microstructured electrode stripes are slightly wedged as shown in Fig.~\ref{fig:trap}(c). This produces an additional component of the electric field $E_z$ parallel to the stripes, thereby eliminating the electric field zero. To avoid ``leaking'' of the molecules from the trap, the microstructured electrodes with the same polarity as the perimeter electrode are interconnected under the perimeter electrode (Fig.~\ref{fig:trap}(d)), causing the holes to lead back into the trap.

\begin{figure}[t]
\centering
\includegraphics[width=0.48\textwidth]{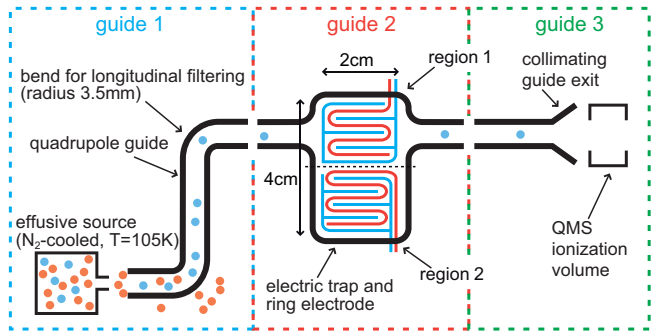}
\caption{(color online). 
Schematic of the setup. The slowest molecules from a liquid-nitrogen-cooled reservoir are loaded into the electric trap via a quadrupole guide connected to the trap. For detection, an exit quadrupole guides the molecules to a quadrupole mass spectrometer (QMS).}
\label{fig:setup}
\end{figure}

Operating the trap requires a suitable source of molecules and a means for their detection. This leads to an integration of the trap in the experimental setup as shown in Fig.~\ref{fig:setup}. As a source of molecules we employ velocity filtering with an electric quadrupole guide from a liquid-nitrogen-cooled effusive nozzle as described in detail elsewhere~\cite{Junglen04}. This method has the advantage of providing a large continuous flux of molecules using a very robust setup. The geometry of the trap is specifically chosen to permit the connection to a quadrupole guide. Here, interrupting the perimeter electrode of the trap allows two opposing electrodes of the quadrupole guide to be connected to the trap. The other two electrodes of equal polarity merge with the microstructured plates. After trapping, the molecules are guided to the ionization volume of a quadrupole mass spectrometer (QMS). This enables time-resolved detection with a simple, generally applicable technique. For signal enhancement, the guide electrodes at the exit of the guide are bent outwards which, similar to a microwave horn antenna, collimates the molecules onto the ionization volume of the QMS. Note that the guide used in our experiment consists of three independently switchable segments, allowing the two outlets of the trap to be electrically closed. 

The measurements are carried out with fluoromethane (${\rm CH_3F}$), a lightweight symmetric-top molecule, but in principle all molecules with significant Stark shifts can be used. The density of trapped ${\rm CH_3F}$-molecules for the maximum trapping fields is approximately $10^8\,\rm{cm^{-3}}$, as has been determined via a QMS calibration~\cite{Sommer09}. This value reflects the density of molecules in the source.

\begin{figure}[t]
\centering
\includegraphics[width=0.45\textwidth]{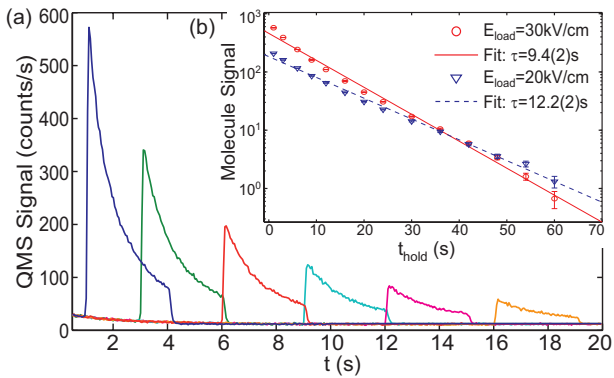}
\caption{(color online). 
(a) Trap unloading signal for $E_{\rm load}=30\,\rm{\frac{kV}{cm}}$ and different holding times $t_{\rm hold}$ versus time $t$ after closing the trap. (b) Integrated unloading signal of the molecules as a function of $t_{\rm hold}$ for two loading field strengths ($20\,\rm{\frac{kV}{cm}}$ and $30\,\rm{\frac{kV}{cm}}$). The blue (dashed) and the red (solid) line are exponential fits for the determination of the lifetime.}
\label{fig:storage}
\end{figure}

For trap characterization, we first determined the trap lifetime by varying the holding time for molecules. 
Initially, molecules are continuously loaded until a steady state is established in the trap. 
This loading process is carried out at reduced trapping and guiding fields $E_{\rm load}$, resulting in a colder molecule ensemble.
Measurements were performed for two different loading fields, as detailed in Fig.~\ref{fig:storage}(b), corresponding to slower and faster molecules~\cite{Junglen04}. This allows us to analyze the dependence of the trap lifetime on the initial velocity distribution of the molecules. After the loading process, the trapping fields are increased to confine the molecules in the trap during the holding time, $t_{\rm hold}$, ranging from $1$ to $60\,\rm{s}$. Simultaneously, high negative voltages are applied to the $1^{\rm st}$ and $3^{\rm rd}$ guide. This creates a repelling electric field at the gaps between the guides and electrically closes the two outlets of the trap. After $t_{\rm hold}$, the trap and the $3^{\rm rd}$ guide are switched back to a guiding configuration with $E_{\rm unload}=E_{\rm load}$ to efficiently extract the molecules from the trap.

Fig.~\ref{fig:storage}(a) shows typical time-of-flight (TOF) signals for the unloading process for different holding times. In Fig.~\ref{fig:storage}(b) the integrated molecule signal for the two different loading fields is plotted as a function of $t_{\rm hold}$. As can be seen, even after $t_{\rm hold}=60\,\rm{s}$ we still measure molecules from the trap. To determine the trap lifetime for slower ($E_{\rm load}=20\,\rm{\frac{kV}{cm}}$) and faster ($E_{\rm load}=30\,\rm{\frac{kV}{cm}}$) molecules, the data are fitted with an exponential decay function. Evidently, slower molecules have a longer trap lifetime which is consistent with Majorana flips being one of the main loss mechanisms for molecules in the trap. Additional contributions might be due to collisions with the background gas (the pressure is $\sim1\times10^{-10}$\,mbar in the trap chamber) or remaining holes in the trap. Lastly, note that the data show slight deviations from the exponential decay function. This is due to a larger initial decay rate which is again consistent with faster molecules getting lost from the trap at a higher rate.

As a second test, we demonstrate the versatility of our trap by performing adiabatic cooling of the molecules. Here, the temperature is reduced by adiabatically expanding a molecular gas from one to both trap regions. This doubling of the trap volume is implemented by ramping down a potential step in the middle of the trap. After loading of slow molecules all voltages are ramped up and a high electric offset field is applied between the plates in region~1, creating a large potential step in the trap. Due to the large voltages, the confinement field between one of the plates in region~1 and the perimeter electrode is zero, causing all molecules not confined to region~2 to be lost from the trap. Next, the offset field in region~1 is ramped down to the offset field in region~2 in the ramping time $t_{\rm ramp}$, thereby doubling the trap volume. This expansion process is expected to conserve the phase-space density of the molecules if it is done adiabatically. Therefore, in the experiment $t_{\rm ramp}$ is varied to analyze the time scale of adiabaticity; a subsequent holding time before unloading is chosen such that $t_{\rm ramp}+t_{\rm hold}=1.1\,\rm{s}=\rm{const}$. 

\begin{figure}[t]
\centering
\includegraphics[width=0.45\textwidth]{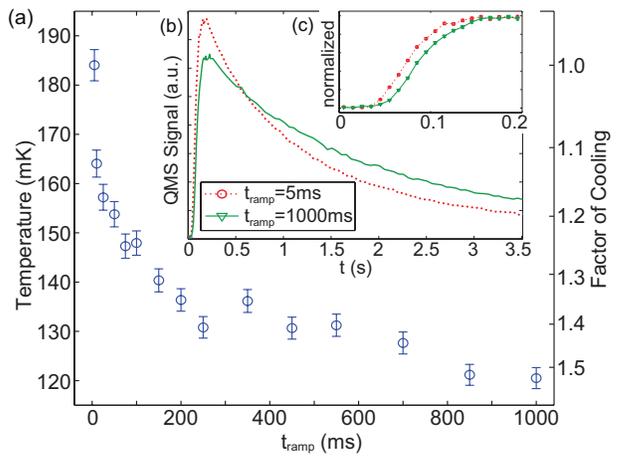}
\caption{(color online). 
(a) Molecule temperature and cooling factor for the adiabatic cooling versus the ramping time. (b) Typical TOF signal. Molecules with $t_{\rm ramp}=1000\,\rm{ms}$ arrive later and decay slower than molecules with $t_{\rm ramp}=5\,\rm{ms}$. (c) Close-up of the normalized rising edge signal.}
\label{fig:adiabatic}
\end{figure}

Figs.~\ref{fig:adiabatic}(b) and (c) compare the TOF unloading signal for the slowest ($t_{\rm ramp}=1000\,\rm{ms}$) and fastest ($t_{\rm ramp}=5\,\rm{ms}$) ramping where the most significant signal difference is expected. As can be seen, for slower ramping of the electric fields the molecules arrive later at the QMS, demonstrating a slower velocity distribution. This is corroborated by the slower decay for the $1000\,\rm{ms}$ ramp since slower molecules have a lower trap loss rate as shown by the trap lifetime measurement. The overall number of measured molecules is even slightly higher for $t_{\rm ramp}=1000\,\rm{ms}$ than for $t_{\rm ramp}=5\,\rm{ms}$. This is clear evidence that the velocity reduction is not due to a filtering process but rather that a new, shifted velocity distribution is created by the ramping. 

We estimate the molecular temperature $T$ for the different ramping times according to $k_{\rm B}T/2=m\left<v_z\right>^2/2$ from the rising edge of the normalized TOF signal $S(t)$. Here, $\left<v_z\right>$ is the mean of the longitudinal velocity distribution $\rho(v_z){\rm d}v_z$ in the exit guide which determines the normalized TOF signal. Using $S(t)=\int_{L/t}^\infty \rho(v_z){\rm d} v_z$ with $L$ being the length of the third guide, we find 

\[
\left<v_z\right>=L\int_0^{\infty} \frac{1}{t^2}S(t)\,{\rm d}t.
\]

In addition to the temperature $T$, we define the cooling factor $F$ for each ramping time as the ratio in $T$ between the given ramping time and the fastest ramping time. The resulting values of $T$ and $F$ are shown in Fig.~\ref{fig:adiabatic}(a) as a function of the ramping time. As expected for a transition from a non-adiabatic to an adiabatic process, we see a steep initial increase of the cooling factor followed by a plateau. The transition between the two at $t_{\rm ramp}\approx100\,\rm{ms}$ corresponds to a molecule with a typical velocity of $6\,\rm{m/s}$ traveling back and forth the full $4\,\rm{cm}$ length of the trap a total of maximally $8$ times. Given the need for a molecule to switch regions several times for the process to be adiabatic, the frequent change in direction of a molecule upon reflection from the microstructure field and the need for the various velocity components to mix, this $100\,\rm{ms}$ timescale of adiabaticity therefore seems reasonable. For $t_{\rm ramp}=1000\,\rm{ms}$ we determine a maximum cooling factor $F_{\rm max}=1.53\pm0.03$, with the corresponding minimal temperature $T_{\rm min}=121\pm2\,\rm{mK}$.

To estimate the yield of the adiabatic cooling we compare the experimental results with the maximum cooling factor we expect from theory. At first, molecules are only confined in trap region~2. When the initial kinetic energy of the molecules exceeds the potential barrier due to the high electric fields in region~1, the molecules can enter this region where they lose energy due to the potential step. If this expansion of the molecular gas to double its volume is done adiabatically, the phase-space density is conserved and the molecular temperature is reduced by a theoretical factor of $F_{\rm opt}=2^{2/d}$, with $d=3$ being the spatial degree of freedom of the contributing velocities. Comparing this theoretically expected maximum cooling factor $F_{\rm opt}=1.59$ to the experimentally measured value results in an experimental yield of ${\rm log}\left(F_{\rm max}\right)/{\rm log}\left(F_{\rm opt} \right)=92\pm3\%$. The main limitation in the experiment is given by the non-zero ramping time of the fastest ramping which is used as the non-adiabatic reference point $T_{\rm max}$ for all data points. Faster ramping than $5\,\rm{ms}$ could result in the demonstration of even higher yields, but is hard to implement due to technical limitations of our voltage supplies. 

In summary, we have presented the first experimental demonstration of a microstructured box-like electric trap with adjustable homogeneous offset fields. Molecules are stored for up to $60\,\rm{s}$ with a trap lifetime of $12.2\pm0.2\,\rm{s}$ which, to our knowledge, is the longest lifetime shown for an electric trap to date. Additionally, adiabatic cooling has been demonstrated with a cooling factor of up to $1.53\pm0.03$ corresponding to a cooling yield of at least $92\pm3\%$. This controlled microstructure-based manipulation of molecules is a major step towards scalable trapping systems as in atom chip experiments~\cite{atomchips}.

Notwithstanding the excellent performance of the trap, further improvements are possible. For example, non-adiabatic transitions as one of the main loss mechanisms can be suppressed by better tailoring the microstructure electrodes. Besides increasing the electrode voltages the density of molecules in the trap can be enhanced by, e.g., combining the trap with a cryogenic buffer-gas cooled source~\cite{Sommer09,Maxwell05} or via laser-induced accumulation of molecules inside our trap~\cite{Zeppenfeld09}.

The present trap already enables a number of measurements. For example, the addition of suitable microwave and optical fields will allow cooling of both the motional and the internal degrees of freedom of polar molecules~\cite{Zeppenfeld09,Vogelius02,Morigi07}. In combination with state-sensitive detection methods~\cite{Motsch07}, the tunable homogeneous offset fields and long trap lifetime can be used for precision Stark spectroscopy or the investigation of field-controlled collision resonances.

We thank S. Chervenkov for helpful discussions. Support by the Deutsche Forschungsgemeinschaft through the excellence cluster ``Munich Centre for Advanced Photonics'' is acknowledged.

\bibliographystyle{unsrt}

\begin{thebibliography}{10}

\bibitem{DeMille02}
D. DeMille, Phys. Rev. Lett. {\bf 88}, 067901 (2002).

\bibitem{Andre06}
A. Andr{\'e} \emph{et al.}, Nature Phys. {\bf 2}, 636 (2006).


\bibitem{Goral02}
K. G\'oral, L. Santos, and M. Lewenstein, Phys. Rev. Lett. {\bf 88}, 170406 (2002).

\bibitem{Micheli06}
A. Micheli, G.\,\,K. Brennen, and P. Zoller, Nature Phys. {\bf 2}, 341 (2006).

\bibitem{Hinds97}
E.\,\,A. Hinds, Phys. Scr. {\bf T70}, 34 (1997).

\bibitem{Vutha10}
A.\,\,C. Vutha, \emph{et al.}, J. Phys. B. {\bf 43}, 074007 (2010).


\bibitem{Hudson11}
J.\,\,J. Hudson \emph{et al.}, Nature (London) {\bf 473}, 493 (2011).

\bibitem{Krems08} 
R.\,\,V. Krems, Phys. Chem. Chem. Phys. \textbf{10}, 4079 (2008).


\bibitem{Miranda11}
M.\,\,H.\,\,G. Miranda \emph{et al.}, Nature Phys. \textbf{7}, 502 (2011).

\bibitem{Avdeenkov02}
A.\,\,V. Avdeenkov and J.\,\,L. Bohn, Phys. Rev. A \textbf{66}, 052718 (2002).


\bibitem{Tscherbul08}
T.\,\,V. Tscherbul and R.\,\,V.~Krems, J. Chem. Phys. \textbf{129}, 034112 (2008).

\bibitem{Meerakker05a}
S.\,\,Y.\,\,T. van de Meerakker \emph{et al.}, Phys. Rev. Lett. {\bf 94}, 023004 (2005).

\bibitem{Rieger05}
T.~Rieger \emph{et al.}, Phys. Rev. Lett. \textbf{95}, 173002 (2005).

\bibitem{Kleinert07}
J. Kleinert \emph{et al.}, Phys. Rev. Lett. \textbf{99}, 143002 (2007).

\bibitem{Hogan09}
S.\,\,D. Hogan, Ch. Seiler, and F. Merkt, Phys. Rev. Lett. {\bf 103}, 123001 (2009).

\bibitem{Zeppenfeld09}
M. Zeppenfeld \emph{et al.}, Phys.\,Rev.\,A\,\textbf{80},\,041401(R)\,(2009).

\bibitem{Wark92}
S.\,\,J.~Wark and G.\,\,I.~Opat, J. Phys. B \textbf{25}, 4229 (1992).

\bibitem{Schulz04}
S.\,\,H. Schulz \emph{et al.}, Phys. Rev. Lett. \textbf{93}, 020406 (2004).

\bibitem{Kirste09}
M. Kirste \emph{et al.}, Phys. Rev. A {\bf 79}, 051401(R) (2009).


\bibitem{Junglen04}
T. Junglen \emph{et al.}, Eur. Phys. J. D {\bf 31}, 365 (2004).

\bibitem{Sommer09}
C.~Sommer \emph{et al.}, Faraday Discuss. {\bf 142}, 203 (2009).

\bibitem{atomchips}
\emph{Special Issue on Atom Chips}, edited by C. Henkel, J. Schmiedmayer, and Ch. Westbrook, Eur. Phys. J. D {\bf 35} (2005).


\bibitem{Maxwell05}
S.\,\,E. Maxwell \emph{et al.},
Phys. Rev. Lett. {\bf 95}, 173201 (2005).


\bibitem{Vogelius02}
I.\,\,S. Vogelius, L.\,\,B. Madsen, and M. Drewsen, Phys. Rev. Lett. {\bf 89}, 173003 (2002).

\bibitem{Morigi07}
G. Morigi \emph{et al.},
Phys. Rev. Lett. \textbf{99}, 073001 (2007).


\bibitem{Motsch07}
M. Motsch \emph{et al.}, Phys. Rev. A \textbf{76}, 061402(R) (2007).

\end{thebibliography}

\end{document}